\documentclass[a4paper]{jpconf}
\usepackage{graphicx}
\usepackage{subfigure}
\usepackage{grffile}
\usepackage{amsmath}
\usepackage{iopams}

\newcommand{\ket}[1]{{\vert #1\rangle}}

\newcommand{\braket}[2]{\langle#1\vert#2\rangle}
\newcommand{\eval}[3]{\langle#1\vert#2\vert#3\rangle}
\newcommand{\vev}[1]{\langle #1\rangle}

\begin{document}
\title{Photoinduced spin-order destructions in one-dimensional extended
Hubbard model}

\author{Hantao Lu$^1$, Shigetoshi Sota$^2$, Hiroaki Matsueda$^3$,
  Janez Bon\v{c}a$^{4,5}$, and Takami Tohyama$^1$}

\address{$^1$ Yukawa Institute for Theoretical Physics, Kyoto
  University, Kyoto, 606-8502, Japan}
\address{$^2$ Computational Materials Science Research Team, RIKEN
  AICS, Kobe, Hyogo 650-0047, Japan}
\address{$^3$ Sendai National College of Technology, Sendai, 989-3128,
  Japan}
\address{$^4$ Faculty of Mathematics and Physics, University of
  Ljubljana, SI-1000 Ljubljana, Slovenia}
\address{$^5$ J. Stefan Institute, SI-1000 Ljubljana, Slovenia}

\ead{luht@yukawa.kyoto-u.ac.jp}

\begin{abstract}
  By employing the time-dependent Lanczos method, the nonequilibrium
  process of the half-filled one-dimensional extended Hubbard model
  under the irradiation of a transient laser pulse is investigated. We
  show that in the spin-density-wave (SDW) phase, the
  antiferromagnetic spin correlations are impaired by the photoinduced
  charge carriers. Near the phase boundary between the SDW and
  charge-density-wave (CDW) phases, a local enhancement of charge
  (spin) order that is absent in the original SDW (CDW) phase can be
  realized with proper laser frequency and strength. The possibility
  of restoration of spin orders from the CDW phase by optical means is
  discussed.
\end{abstract}

\section{Introduction}

New insights into the dynamics of strongly correlated electron systems
can be achieved by investigating their nonequilibrium processes, which
demands the knowledge and understanding of the properties of excited
states. The process have to be treated nonperturbatively if the system
is driven far from the equilibrium.

In experiments, one way to achieve nonequilibrium states is by
applying various external electric fields, including pulses of
electromagnetic field, which couple with charge degrees of freedom of
the system. Well-known experimental facts on (quasi) one-dimensional
(1D) materials include dielectric
breakdown~\cite{Tokura:1988,Taguchi:2000}, insulator to metal
transitions and so on~\cite{Iwai:2003,Okamoto:2007kc,Kimura:2009by}.

In Mott insulators, we have to deal with the states above the Mott gap
in order to induce charge carriers optically. It is well known that
the ground states of Mott insulators have quasi-long-range
antiferromagnetic spin orders. The charge carriers created by photons
are, thus, expected to affect the spin correlations. It is desirable
to obtain the knowledge of spin dynamics, as well as the charge's,
during and after the doping. In many 1D materials, it has been shown
that besides the on-site Coulomb interaction $U$, nearest-neighbor
(NN) interaction $V$ has to be taken into
account~\cite{Yamashita:1999bt,kumar:234304}. With increase of $V$, a
phase transition from the Mott insulator to a charge-order phase can
occur. The nonequilibrium dynamics near the phase boundary is also
important.

In this proceeding, we investigate the nonequilibrium process of a 1D
strongly correlated system under the irradiation of a laser pulse. The
model we are working on is the half-filled extended Hubbard model,
where the on-site, and NN interactions ($U$ and $V$), are
included. Its phase diagram in equilibrium is well-known and
understood
~\cite{vanDongen:1994ca,Nakamura:2000gk,Tsuchiizu:2002eg,Ejima:2007go}. Here
we would like to reiterate some known properties of the model,
concerning our work. In strong-coupling regime (large $U(>0)$), with
increase of $V(>0)$, a first-order phase transition from the
spin-density-wave (SDW) to charge-density-wave (CDW) phase is realized
when $V$ reaches around $U/2$. The transition can be understood as
driven by the competition between the energetic costs for doublon
generation and the benefits due to the attraction between doublons and
holons. The SDW and CDW phases are recognized by algebraic decay of
spin correlations, and a long-range (staggered) charge order,
respectively. In both phases, the charge sector is always gapped
except on the boundary. When spin degrees of freedom are concerned,
gapless spin modes can be found in the SDW phase, while absent in the
CDW phase.

In the following sections, by studying time-dependent spin and charge
correlation functions numerically, we show that spin correlations are
suppressed by the photoinduced charge carriers in the SDW phase. Near
the phase boundary from the SDW side, with proper laser frequency and
strength, a sustainable charge order enhancement can be
realized~\cite{Lu:2012ui} but local spin correlations
remains. Analogously, from the CDW side, the suppression of long-range
charge order is accompanied with a local spin correlation enhancement.

\section{Model and Numerical Method}

The time-dependent Hamiltonian for the 1D extended Hubbard model,
where the external field is incorporated by means of the Peierls
substitution for the hopping coefficients, can be written as
\begin{eqnarray}
H(t)&=&-t_h\sum_{i,\sigma}\left(e^{iA(t)}c_{i,\sigma}^{\dagger}
c_{i+1,\sigma}+\text{H.c.}\right) \nonumber \\
&&+U\sum_{i}\left(n_{i,\uparrow}-\frac{1}{2}\right)
\left(n_{i,\downarrow}-\frac{1}{2}\right) 
+V\sum_i\left(n_i-1\right)\left(n_{i+1}-1\right),
\label{eq:1}
\end{eqnarray}
where $c_{i,\sigma}^\dagger$ ($c_{i,\sigma}$) creates (annihilates)
electrons with spin $\sigma$ at site $i$,
$n_{i,\sigma}=c_{i,\sigma}^\dagger c_{i,\sigma}$,
$n_i=n_{i,\uparrow}+n_{i,\downarrow}$, $t_h$ is the hopping
constant. Particularly, with temporal gauge, we use a time-dependent
vector potential $A(t)$ to describe the laser
pulse~\cite{Matsueda:2010ue}
\begin{equation}
A(t)=A_0e^{-\left(t-t_0\right)^2/2t_d^2}\cos
\left[\omega_{\text{pump}}\left(t-t_0\right)\right],
\label{eq:2}
\end{equation}
where $A_0$ controls the laser amplitude, which reaches its full
strength at $t=t_0$; $t_d$ characterizes the duration time of light
action. Note that, due to finite $t_d$, the incoming photon frequency
is broadened into a Gaussian-like distribution, with the variance of
$1/t_d^2$ around the central value $\omega_{\text{pump}}$. Throughout
this paper, we set $t_h$ and $t_h^{-1}$ as energy and time units.

Starting from the Schr\"odinger equation $i\partial\psi(t)/\partial
t=H(t)\psi(t)$, in order to obtain $\psi(t)$, we employ the
time-dependent Lanczos method, which is originally described in
Ref.~\cite{park:5870}, and followed by its applications in
nonequilibrium dynamics of strongly correlated systems recently in
Ref.~\cite{Prelovsek2011.5931p}. The basic idea is that we approximate
the time evolution of $\ket{\psi(t)}$ by a step-vise change of time
$t$ in small increments $\delta t$. At each step, the Lanczos basis
with dimension $M$ is generated resulting in the time evolution
\begin{equation}
\ket{\psi(t+\delta t)}\simeq e^{-iH(t)\delta t}\ket{\psi(t)}
\simeq\sum_{l=1}^M e^{-i\epsilon_l\delta t}\ket{\phi_l}\braket{\phi_l}{\psi(t)},
\label{eq:3}
\end{equation}
where $\epsilon_l$ and $\ket{\phi_l}$, respectively, are eigenvalues
and eigenvectors of the tridiagonal matrix generated in Lanczos
iteration. As one of the advantages, the Lanczos method preserves the
unitarity of the time-evolution operator at each time step, with
relative low cost for a desired accuracy. The convergence check shows
that for time length $t$ around hundred, taking $M=30$ with $\delta
t\lesssim 0.1$ can provide adequate accuracy for the parameter regime
we are working in. For even smaller $\delta t$, $M$ can be smaller.

In the succeeding numerical calculations, periodic boundary conditions
are employed to avoid possible boundary effects. We fix the parameters
$t_d=5$, $t_0=12.5$, and chose $U=10$, where the first-order
transition occurs around $V\approx 5.1$~\cite{Ejima:2007go}.

\section{Numerical Results}

In order to investigate the time evolutions of the spin and charge
orders under the pules, we define the spin-spin and charge-charge
correlations, respectively:
\begin{subequations}
\begin{equation}
S(j;t)=\frac{(-1)^j}{L}\sum_{i=0}^{L-1}
\eval{\psi(t)}{{\bf S}_{i+j}\cdot{{\bf S}_{i}}}{\psi(t)},
\label{eq:4}
\end{equation}
\begin{equation}
C(j;t)=\frac{(-1)^j}{L}\sum_{i=0}^{L-1}
\eval{\psi(t)}{(n_{i+j}-1)(n_i-1)}{\psi(t)},
\label{eq:5}
\end{equation}
\end{subequations}
where $L$ is the lattice size, and ${\bf S}_i$ is the spin operator on
site $i$. Intuitively, the antiferromagnetic spin and staggered charge
orders are competitive. 

\begin{figure}
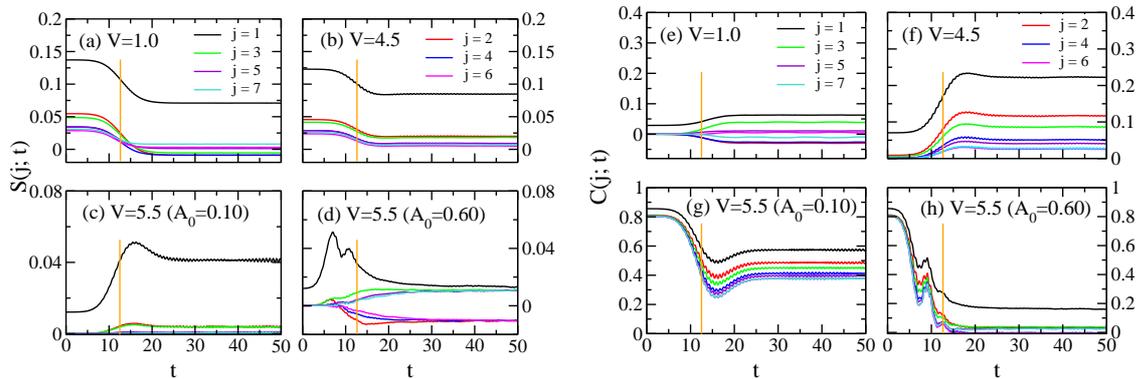

\begin{tabular}{cc}
\includegraphics[width=0.45\textwidth]{fig1a} &
\includegraphics[width=0.45\textwidth]{fig1b} \\
\end{tabular}
\caption{Time dependence of the spin-spin (left, from (a) to (d)) and
  charge-charge (right, from (e) to (h)) correlations as functions of
  distance (labeled by $j$) for $14$-site lattice. The laser pulse
  with Gaussian magnitude modulation reaches its full strength at
  $t=12.5$, as indicated by solid lines. Without exception, the
  pumping frequencies are set to match the resonance peaks of the
  optical absorption spectra. Parameters: $\delta t=0.02$, $M=100$. In
  each subfigure, (a),(e): $V=1.0$, $\omega_{\text{pump}}=7.1$,
  $A_0=0.10$; (b),(f): $V=4.5$, $\omega_{\text{pump}}=4.0$,
  $A_0=0.07$; (c),(g): $V=5.5$, $\omega_{\text{pump}}=4.1$,
  $A_0=0.10$; (d),(h): the same as (c) (or (g)), except for
  $A_0=0.60$.}
\label{fig:1}
\end{figure}

Figure~\ref{fig:1} shows the results of correlations $S(j;t)$,
$C(j;t)$, up to $t=50$, for the system with $14$ lattice sites, where
the largest distance between two sites is $7$. Total crystal momentum
$P$, as a good quantum number, is implemented. The size of the Hilbert
space for $P=0$ is 841332. For 2600 time step, it took around 30 hours
for 12 cores. In succeeding discussions, we put more emphasis on
spin. Detailed discussions on the charge correlations can be found in
Ref.~\cite{Lu:2012ui}.

In Fig.~\ref{fig:1}, four cases are shown: (a) and (e), deep into the
SDW phase ($V=1$); (b) and (f), the ground state is near the phase
boundary, but set on the SDW side ($V=4.5$); (c) and (g), (d) and (h),
at the CDW side ($V=5.5$) with different laser strength. Several
observations can be made. First, notice that for the ground states
(corresponding to $t=0$) in the SDW phase, we can observe the
gradually decay of spin correlations (Figs.~\ref{fig:1}(a), (b)) while
no long-range charge order exists (Figs.~\ref{fig:1}(e), (f)); in the
CDW phase, the presence of a long-range charge order can be clearly
recognized (Figs.~\ref{fig:1}(g) or (h)), and the spin correlations
largely vanish except for the NN one (See Figs.~\ref{fig:1}(c) or (d),
where $S(j;0)\lesssim 10^{-4}$ for $j>1$, and $S(1;0)\sim
0.01$). Second, for the case of $V=1.0$, where the system is deep
inside the SDW phase, introducing charge carriers into the system by
the pulse cannot produce charge order (Fig.~\ref{fig:1}(e)), while the
spin correlations are severely suppressed except for the NN one (See
Fig.~\ref{fig:1}(a), where $S(1;t)$ decreases from $0.14$ at $t=0$ to
$0.07$ at $t=50$, while some correlations with other distances even
become negative). Notice that in this case, after the pulse, the total
number of doublons (defined as $N_{\text{db}}=\sum_i n_{i,\uparrow}
n_{i,\downarrow}$) is around $2.2$, corresponding to $30\%$
doping~\cite{Takahashi:2008bk}. Third, we would like to pick up
Fig.~\ref{fig:1}(f), which shows that proper incoming laser pulse can
enhance the charge order considerably in the SDW near the phase
boundary, where we can observe that the NN correlation increases from
$0.07$ at $t=0$ to $0.22$ at $t=50$, together with a clear enhancement
of other correlations in larger distances, which start from almost
zero magnitudes at $t=0$; meanwhile, the spin order does not suffer
heavy suppression (See Fig.~\ref{fig:1}(b), where the NN correlation
drops from $0.12$ to $0.08$). Analogously, from the CDW side close to
the boundary, the pulse can suppress the charge order
(Fig.~\ref{fig:1}(g)) with local spin enhancement taking place at the
same time (Fig.~\ref{fig:1}(c), where the NN correlation increases
from $0.01$ to $0.04$). Finally, from Fig.~\ref{fig:1}(d) and (h), we
show that strong enough laser pulse can destroy the charge order
substantially without spin correlation enhancement.

\begin{figure}
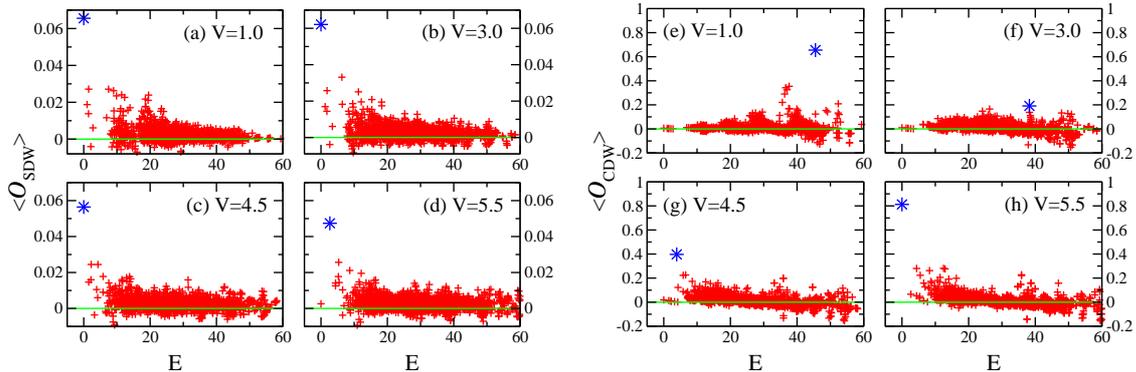

\begin{tabular}{cc}
\includegraphics[width=0.45\textwidth]{fig2a} &
\includegraphics[width=0.45\textwidth]{fig2b} \\
\end{tabular}
\caption{The expectations of SDW (from (a) to (d)) and CDW (from (e)
  to (h)) order parameters of eigenstates for $10$-site lattice with
  $V=1.0$, $3.0$, $4.5$, and $5.5$. The energy $E$ is measured from
  the ground state. Only the data of states with
  $P_{\text{tot}}=S_{\text{tot}}=0$ are shown, up to $E=60$. Note that
  the largest values of the order parameters in (a) through (h) are
  marked by stars.}
\label{fig:2}
\end{figure}

The results of the time-dependent correlations suggest intimate
connections between spin and charge degrees of freedom in the
photoexcited states. Interesting phenomena occur around the phase
boundary, where local spin and charge orders can coexist.  For a more
quantitative analysis, we compare the spectra of systems with
different values of $V$ in the time-independent Hamiltonian, i.e.,
Eq.~(\ref{eq:1}) but $A(t)=0$. In order to describe the spin and
charge orders which have finite spatial extensions, we define
\begin{subequations}
\begin{equation}
\mathcal{O}_{\text{SDW}}=\frac{1}{LL_c}\sum_{i=0}^{L-1}\sum_{j=1}^{L_c}
(-1)^j{\bf S}_{i+j}\cdot{\bf S}_i
\label{eq:6}
\end{equation}
\begin{equation}
\mathcal{O}_{\text{CDW}}=\frac{1}{LL_c}\sum_{i=0}^{L-1}\sum_{j=1}^{L_c}
(-1)^j\left(n_{i+j}-1\right)\left(n_i-1\right).
\label{eq:7}
\end{equation}
\end{subequations}
Here $L_c$ is introduced as a cut off parameter for the correlation
length, and we set $L_c=5$, estimated from the results of the
density-matrix renormalization group (DMRG) on larger system size up
to $30$. The expectation values of the order parameter
$\vev{\mathcal{O}_{\text{CDW}}}$ and $\vev{\mathcal{O}_{\text{SDW}}}$
for eigenstates of a smaller 10-site system are shown in
Fig.~\ref{fig:2}.

In Fig.~\ref{fig:2} from (a) to (d), we can notice that for $V=1.0$ to
$5.5$, states with relative large $\vev{\mathcal{O}_{\text{SDW}}}$ are
always situated in the low-energy regime (say, $E<10$). On the
contrary, states with prominent $\vev{\mathcal{O}_{\text{CDW}}}$ only
appear there when the system is approaching the phase boundary. In
Fig.~\ref{fig:2}(g) for $V=4.5$, around the resonance frequency of
optical absorption spectrum ($\omega_R\approx 4.3$), states with
noticeable values of charge order can be found. This explains why a
considerable charge order enhancement can be observed when the
frequency of laser with proper strength is tuned to $\omega_R$ (see
Fig.~\ref{fig:1}(f), where $\omega_R\approx 4.0$ for $14$-site
lattice). However, in the complementary case, only the enhancement of
the NN spin correlation is apparently visible, with rapid decrease
with increase of distance. The same holds for $L=10$, in spite of the
fact that a state with excitation energy $E\approx 2.7$ has a
distinguished $\vev{\mathcal{O}_{\text{SDW}}}$ (shown by star in
Fig.~\ref{fig:2}(d)).

\begin{figure}
\begin{tabular}{ccc}
\includegraphics[width=0.3\textwidth]{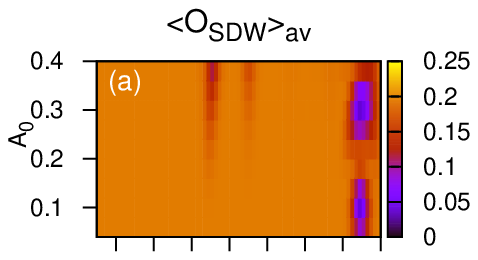} &
\includegraphics[width=0.3\textwidth]{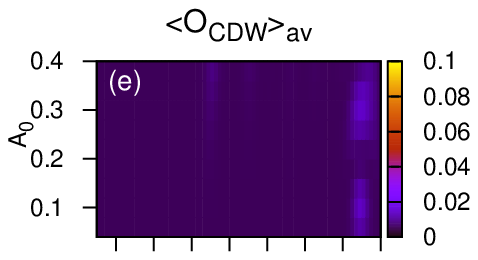} &
\includegraphics[width=0.3\textwidth]{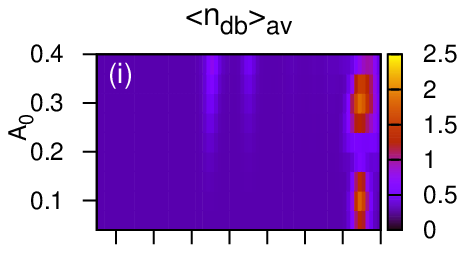} \\
\end{tabular}
\vskip -0.55in
\begin{tabular}{ccc}
\includegraphics[width=0.3\textwidth]{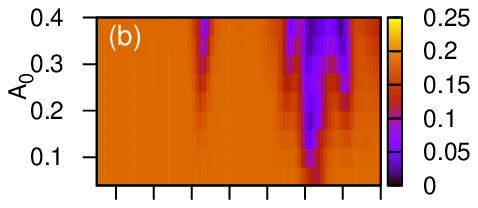} &
\includegraphics[width=0.3\textwidth]{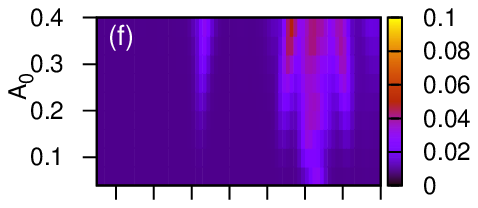} &
\includegraphics[width=0.3\textwidth]{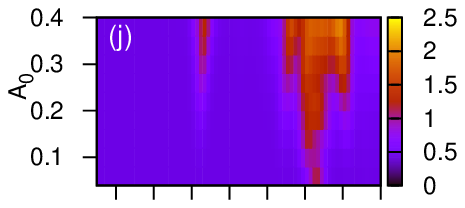} \\
\end{tabular}
\vskip -0.55in
\begin{tabular}{ccc}
\includegraphics[width=0.3\textwidth]{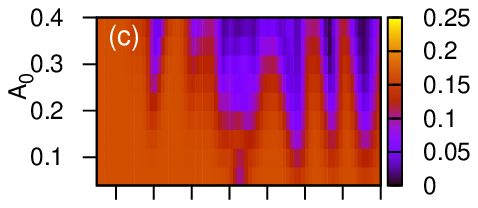} &
\includegraphics[width=0.3\textwidth]{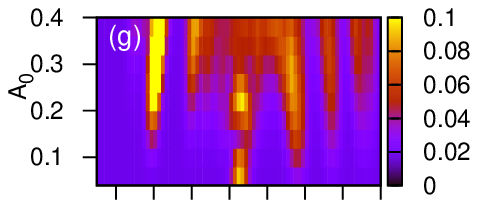} &
\includegraphics[width=0.3\textwidth]{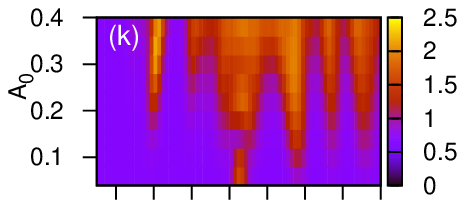} \\
\end{tabular}
\vskip -0.55in
\begin{tabular}{ccc}
\includegraphics[width=0.3\textwidth]{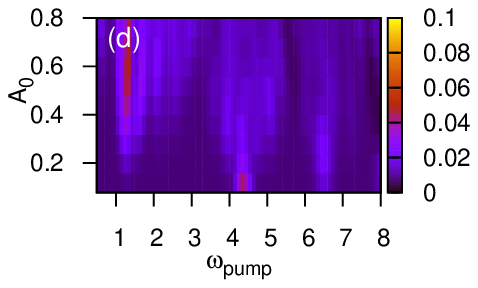} &
\includegraphics[width=0.3\textwidth]{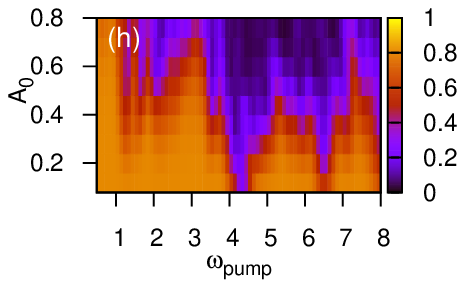} &
\includegraphics[width=0.3\textwidth]{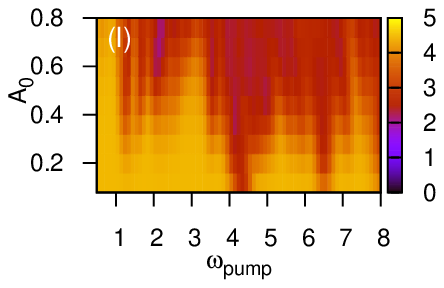} \\
\end{tabular}
\caption{Contour plots of final time-evolution results of SDW order
  $\vev{\mathcal{O}_{\text{SDW}}}_{\text{av}}$ (first column), CDW
  order $\vev{\mathcal{O}_{\text{CDW}}}_{\text{av}}$ (second column)
  and doublon number $N_{\text{db}}$ (last column) as functions of
  $\omega_{\text{pump}}$ and $A_0$, obtained by averaging on the last
  $50$ time steps (time length $\Delta t=5$), for $10$-site
  lattice. $\omega_{\text{pump}}\in[0.5,8]$; $A_0\in[0.04,0.4]$ for
  the first three rows, and $[0.08,0.8]$ for the last row. From top to
  bottom, $V=1.0$, $3.0$, $4.5$, and $5.5$, respectively. Here, we
  take $\delta t=0.1$, $M=30$.}
\label{fig:3}
\end{figure}

In order to understand the different behavior of spin and charge
correlations, and to further elaborate on the condition for the
emergence of order enhancements induced by the laser pulse, we perform
parameter-sweeping calculations in terms of $\omega_{\text{pump}}$ and
$A_0$ for $10$-site lattice. For a given pair of
$\omega_{\text{pump}}$ and $A_0$, we carry out the time evolutions up
to $t=52.5$, then calculate the expectations of the SDW order , CDW
order, and doublon number, noted as
$\vev{\mathcal{O}_{\text{SDW}}}_{\text{av}}$,
$\vev{\mathcal{O}_{\text{CDW}}}_{\text{av}}$, and
$\vev{n_{\text{db}}}_{\text{av}}$, respectively, after averaging on
the last $50$ time steps (corresponding to time length $\Delta
t=5$). The results, presented in contour plots, are shown in
Fig.~\ref{fig:3}.

In Fig.~\ref{fig:3}, we notice that for each $V$, there exist some
frequencies with which the optical absorptions are most efficient,
e.g., $\omega_{\text{pump}}\approx 7.5$ in Fig.~\ref{fig:3}(i) for
$V=1.0$, $\omega_{\text{pump}}\approx 6.3$ in Fig.~\ref{fig:3}(j) for
$V=3.0$. These are resonance frequencies of optical absorption spectra
$\omega_R$, which can be obtained from the imaginary parts of the
dynamical current-current correlation functions. 

We can see that in order to excite the system effectively with weak
fields, we need to tune the pulse frequency $\omega_{\text{pump}}$
close to $\omega_R$. In general, in the SDW phase, accompanied with
photodoping, spin correlations are suppressed (see the correspondence
between the bright regions in Figs.~\ref{fig:3}(i), (j), (k) for
doublon generation, and the dark regions in Figs.~\ref{fig:3}(a), (b),
(c) for $\vev{\mathcal{O}_{\text{SDW}}}_{\text{av}}$). On the other
hand, in the CDW phase, the destruction of charge order alone does not
sufficiently bring back the spin order (see Fig.~\ref{fig:3}(l),
compared with (d)).

Things can be more subtle near the phase boundary. With $A_0$ less
than $0.1$, a local enhancement of CDW order in the SDW phase can be
induced accompanied with a suppression of spin orders, and vice versa
in the CDW phase near the boundary. For example, see around
$\omega_{\text{pump}}\approx 4.3$ in Figs.~\ref{fig:3}(g) and (c) for
the case of $V=4.5$, and $\omega_{\text{pump}}\approx 4.4$ in
Figs.~\ref{fig:3}(d) and (h) for the case of $V=5.5$. Among them, the
weakness of the SDW enhancement in Fig.~\ref{fig:3}(d) can be
attributed to the fact that the states with distinguished
$\vev{\mathcal{O}_{\text{SDW}}}$ (around $E\approx 2.7$, see
Fig.~\ref{fig:2}(d)) are in the off-resonance region, which makes it
hard for them to be excited by the laser.

Going to the high-field region, we observe that in
Fig.~\ref{fig:3}(d), at $\omega_{\text{pump}}\approx 1.2$, there is a
well resolved stripe when $A_0>0.4$, indicating an enhancement of the
spin order. From the results of $S(j;t)$ (not shown here), we notice
that the spin correlations there extend themselves beyond NN. This may
come from nonlinear multiphoton optical
process~\cite{Oka:2011vw}. Similar phenomenon for the case of charge
order, can be found in Fig.~\ref{fig:3}(g), at
$\omega_{\text{pump}}\approx 2$. Further investigation on the
nonlinear effect is in progress.

\section{Conclusions}

In summary, by studying time-dependent spin and charge correlation
functions of the 1D extended Hubbard model with applied laser pulse,
we have shown that in the SDW phase, the spin order is suppressed due
to the optical doping. Starting from the SDW side not far from the
phase boundary, when the incoming laser frequency matches the optical
absorption peak, a sustainable charge order enhancement can be
realized. Similarly, starting from the CDW side, an enhancement of
spin correlations, though more localized, has been observed. In the
off-resonance region, we find more extended recovery of spin
correlations which may come from nonlinear effects.

\section*{References}
\bibliography{photoinduced1D}

\end{document}